# Motion Artifact Reduction in Quantitative Susceptibility Mapping using Deep Neural Network

Chao Li[1,2], Hang Zhang[2,3], Jinwei Zhang[2,4], Pascal Spincemaille[2], Thanh D.Nguyen[2], Yi Wang[2,4]

1. Department of Applied and Engineering Physics, Cornell University, Ithaca, NY, USA

2. Department of Radiology, Weill Medical College of Cornell University, New York, NY, USA

3. Department of Electrical and Computer Engineering, Cornell University, Ithaca, NY, USA

4. Department of Biomedical Engineering, Cornell University, Ithaca, NY, USA

## Abstract

An approach to reduce motion artifacts in Quantitative Susceptibility Mapping using deep learning is proposed. We use an affine motion model with randomly created motion profiles to simulate motion-corrupted QSM images. The simulated QSM image is paired with its motion-free reference to train a neural network using supervised learning. The trained network is tested on unseen simulated motion-corrupted QSM images, in healthy volunteers and in Parkinson's disease patients. The results show that motion artifacts, such as ringing and ghosting, were successfully suppressed.

## 1. Introduction

Patient movement during MRI can cause degradation of image quality, which manifests as ringing, blurring and ghosting artifacts (1, 2). Motion artifacts have an average 2.5% prevalence in clinical practice (3, 4) and higher prevalence in patient cohorts with little motion control. Motion artifact may result recalled scan and errors in subsequent post-processing analysis and diagnosis (4).

Many approaches have been proposed to reduce or correct motion artifacts. These methods can be broadly categorized as prospective and retrospective. Prospective methods measure patient motion and update acquisition parameters during the MRI acquisition process. The motion parameter estimation can be achieved by navigator-based methods or external device-based method. The navigator-based methods include frequently inserting additional MR "navigator" data acquisitions. These inserted navigators are compared with a reference to extract the rotation and translation motion parameters (5-10). The external device-based techniques use external tracking systems to estimate motion. Examples are optical cameras tracking a marker on the patient's body and field probes that measure magnetic field perturbations (11-13). Having the motion trajectories, motion correction can be performed either prospectively or retrospectively. In autofocus motion correction, motion parameters are extracted from the motion-corrupted images themselves, for example by minimizing image entropy (14), by using projections onto convex sets (POCS) (15, 16) or by pMRI-based methods (17). Motion artifacts manifest as corrupted k-space lines, and these corrupted samples are detected and corrected using these methods.

In the past decade, deep learning has demonstrated its potential in many medical imaging applications. Deep neural networks, especially convolutional neural networks (CNNs), are trained to perform tasks like image reconstruction of under-sampled data (18), structural segmentation (19), and medical image translations between different modalities (20). CNNs have also been used to detect motion and correct motion artifacts. A 3D-based conditional generative adversarial network has been used to correct simulated and real-world artifacts in gradient echo images (20). The authors in reference (13) trained a CNN to remove artifacts from images in order to improve a model-based motion estimation. The CNN -estimated motion-corrected images served as image priors for the motion parameter search. In the work by Wang et al. (21), a pretrained CNN again is used to generate an image prior to compensate the motion artifact in the FOV due to the motion out-of-FOV with acquired motion parameters from tracking cameras, navigators, etc.

Quantitative susceptibility mapping (QSM) (22) is an MRI technique that is able to measure magnetic susceptibility in vivo, allowing the measurement of biomarkers such as iron, calcium and gadolinium for assessing pathologies. It has been applied to study paramagnetic iron and gadolinium and diamagnetic calcification and myelin in health and diseases (22-30). Mathematically, QSM solves am ill-posed inverse problem. A breakthrough for QSM was coined in 2010 using the Bayesian maximum a posteriori (MAP) method (31). Recently, QSM has demonstrated its ability in assessing Parkinson disease (PD) as there is an increased iron deposition in the substantia nigra of patients with idiopathic Parkinson's disease (32). Similar to other MRI techniques, motion artifacts can be seen in QSM, particularly when scanning Parkinson's disease (PD) patients who have action tremors which is a common symptom of PD.

In this work, we propose a method to simulate motion artifact in QSM. The simulated motion-corrupted QSM and corresponding motion-free images are used to train a deep convolutional neural network for motion correction. The neural network is tested on unseen QSM images with simulated motion artifacts and real-world motion-corrupted QSM images from PD patients and healthy subjects. To the best of our knowledge, this is the first work on motion artifact compensation in QSM using deep learning techniques.

## 2. Theory

### 2.1. Quantitative susceptibility mapping

Quantitative susceptibility mapping is derived from the phase of multi-echo gradient echo (MGRE) MRI data. The MGRE phase is equal to the magnetic field multiplied by the gyromagnetic ratio $\gamma$ and the echo time (TE). The MGRE signal detected in a voxel centered at $r$ at time $TE$ can be expressed as

$$S(r,TE) = m(r)e^{-\frac{TE}{T2^*}}e^{-ib(r)\omega_0 TE}, \qquad (1)$$

where $m(r)$ contains proton density, T1 weighting and other magnitude contrast, $\omega_0 = \gamma B_0$ is the frequency of proton in a main magnetic field $B_0$ and $b(r)$ is the total field inhomogeneity at point $r$(22).Followed by a phase unwrapping and background field removal, the local susceptibility can be obtained by solving a susceptibility-magnetic field inverse problem,

$$b(r) = (d * \chi)(r) \qquad (2)$$

where $d$ is dipole kernel.

However, the inversion of field to susceptibility is ill-posed as there are zeros in the dipole kernel (33) and can cause severe streaking artifacts in the reconstructed susceptibility map(34, 35). Fortunately, MRI information on tissue anatomical structure can serve as a prior in Bayesian regularization to find a unique solution of the susceptibility, which is called Morphology Enabled Dipole Inversion (MEDI) (36):

$$\hat{x} = \underset{\chi}{\mathrm{argmin}} \frac{1}{2} \|W(d * \chi - b)\|_2^2 + \lambda \|M_G \nabla \chi\|_1 \tag{3}$$

with $\chi$ the susceptibility distribution to be solved, $b$ the field measurement, $d$ the dipole kernel, and $W$ an acquisition noise weighting. The regularization (second term) is a weighted total variation, with $\nabla$ the gradient operator, $M_G$ a binary edge mask determined from the magnitude image(36) which enforces morphological consistency between magnitude and susceptibility, and $\lambda$ the regularization parameter.

## 2.2. Motion artifact simulation in QSM

Patient motion during MR data acquisition leads to artifacts in the reconstructed image. In this work, we assume the motion within each TR can be neglected. Motion leads the combination of k-space corresponding to different head (position) states. This motion corrupted k-space can be considered as a linear combination of k-space segments :

$$S_m(r, TE) = IFFT\left\{\sum_i M_i \odot FFT\left[m(A_i \cdot r)e^{-\frac{TE}{T2^*}}e^{-id*[\chi(A_i \cdot r)]\omega_0 TE}\right]\right\} \tag{4}$$

where $A_i$ is the affine matrix acting on the coordinate of the image at motion state $i$, $\odot$ denotes pointwise multiplication and $M_i$ is the binary mask indicating the k-space acquired during motion state $i$. The phase factor in Eq.(4) is written as an affine transformation acting on the coordinate of the susceptibility source which is then convolved with the dipole kernel. Normally, if the affine transformation contains a rotation, the affine transformation and dipole convolution do not commute, i.e. the order of these two operations cannot be exchanged, i.e. $d * [\chi(A_i \cdot r)] \neq [d * \chi](A_i \cdot r) = b(A_i \cdot r)$. If motion $A_i$ is purely translational, then these two operations do commute and it can be written as $d * [\chi(A_i \cdot r)] = [d * \chi](A_i \cdot r) = b(A_i \cdot r)$. Strictly speaking, the susceptibility source must be calculated first using the original multi-echo GRE in order to generate the motion simulation using Eq.(4). A cohort study found that motion occurs during MRI acquisition is normally in the range of $\pm 5mm$ and $\pm 5°$(37). Within this range, we can approximate $d * [\chi(A_i \cdot r)]$ by $[d * \chi](A_i \cdot r) = b(A_i \cdot r)$, i.e. exchanging the order of affine transformation and dipole convolution, to get the field on the phase for a good approximation. An example is shown in Figure 1, where $d * [\chi(A_i \cdot r)] \approx [d * \chi](A_i \cdot r)$ by a $5°$ rotation ($A_i$ is rotation by 5 degree clockwise). Therefore, Eq.(4) is modified as

$$S_m(r, TE) \approx IFFT\left\{\sum_i M_i \odot FFT\left[m(A_i \cdot r)e^{-\frac{TE}{T2^*}}e^{-ib(A_i \cdot r)\omega_0 TE}\right]\right\} \tag{5}$$

and the imaginary and real part of the original multi-echo GRE signal can be transformed in the same way to simulate the motion-corrupted signal.

## 3. Methods

In this paper, we use the proposed simulation method to simulate motion-corrupted QSM from motion-free multi-echo GRE (MGRE) data. The resulting motion-corrupted and original motion-free image pairs are used as a training example to train a neural network. The trained network is then tested on unseen simulated images and in vivo datasets from Parkinson's disease patients and healthy subjects with real motion artifacts.

### 3.1 Data acquisition and preprocessing

#### 3.1.1 Motion-free dataset for motion artifact simulation

MRI was performed on 80 patients using a 3T system (GE, Signa HDxt) with a 3D MGRE sequence. Detailed imaging parameters included FA = 15-20°, TE1 = 3.5-6.3 ms, TR = 48ms-58ms, #TE = 11, $\Delta$TE = 3.8-5.5ms, matrix size= 512×512×48-60, voxel size = 0.4688×0.4688×3$mm^3$. 70 of these 80 patient images were selected for further processing, with 10 images with degraded image quality, for example significant blurriness, and outlier acquisition protocol parameters were excluded. This dataset only contained motion-free images and was used to simulate motion corrupted QSM for training, validating and testing the neural network. All the images are interpolated to a voxel size of 0.75×0.75×3 $mm^3$ in order to give the neural network a larger receptive field for a fixed patch size.

#### 3.1.2 Dataset with real motion artifact

A second dataset was obtained by performing MGRE MRI on 5 Parkinson's patients that contained real motion artifacts. This data was acquired using a 3T system (Magnetom Skyra, Siemens Healthcare, Erlangen, Germany) and imaging parameters included FA = 15°, FOV = 24.0 cm, TE1 = 6.69 ms, TR = 44 ms, #TE = 10, $\Delta$TE = 3.6 ms, matrix size= 260×320×128 voxel size = 0.8×0.8×0.5 mm3. The reconstructed QSM images were interpolated to the same voxel size as the training data, 0.75×0.75×3 $mm^3$.

A third dataset was collected from 4 healthy subjects using a 3T system (GE, Discovery MR750) with a 3D MGRE sequence. Detailed imaging parameters included FA = 15°, TE1 = 6.3 ms, TR = 49 ms, #TE = 11, $\Delta$TE = 4.1 ms, matrix size= 512×512×60, voxel size = 0.4688×0.4688×3$mm^3$. Each healthy subject was scanned twice. In the first scan, the subject was asked to stay still, which gives a reference image without motion. In the second scan, each subject was allowed to make some movement twice or three times during the scan, each movement lasts roughly 10 to 20 seconds. The reconstructed QSM images were again interpolated to the voxel size 0.75×0.75×3 $mm^3$.

### 3.2 Motion artifact simulation

For each MGRE volume, a sequence of new MGRE volumes corresponding to different head states is simulated using Eq. 5 based on a specific motion trajectory using translation and rotation transformation. For each head state, the same affine transformation is performed on the magnitude and phase of the MGRE data according, which are Fourier-transformed to k-space. Then, the simulated k-space volumes are masked and combined to get the motion-corrupted k-space. This k-space volume is transformed back to image space using inverse Fourier transform, leading to the final motion-corrupted MGRE data. Following that, the local tissue field was estimated using non-linear fitting across multi-echo phase data (38) followed by graph-cut based phase unwrapping (39) and background field removal (40) to obtain the motion corrupted QSM image.

Rigid body motion was simulated for a Cartesian trajectory following the routine described in (37). However, in our work, motion profiles were generated in the range of ±5mm for translation and ±5° for rotation. It is realistic that patient motion only occurs occasionally during the scanning but has significant effects on the image quality, thus we assume that each motion profile contains 2 or 3 events. The motion profile is defined by the number of events, time of onset of each event, and magnitude of the 6 rigid-body motion parameters. 10 motions were simulated for each motion-free GRE data, giving a total of 700 motion corrupted multi-echo GRE data for network training and evaluation.

### 3.3 Network structure

We implemented a 3D U-Net (41) followed by a denoiser. The 3D U-Net was a fully convolutional network architecture backboned with a VGG network. The convolutional kernel size was 3×3×3. With the skip-connect structure in U-Net, low-level features of the encoder can be passed to the decoder, allowing a strong correlation between the input image and the output image, which is suitable for most motion reduction tasks where motion artifact does not erase major structural information of the image. The denoiser was a 2D progressive recurrent network (PReNet). It is used to refine the results output by the U-Net. The PReNet has four parts: (1) a convolution layer that receives the network input, (2) a recurrent network using Long Short-Term Memory (LSTM), (3) several residual blocks that extract features of different levels, (4) a convolutional layer that outputs the denoised image.

The 3D U-Net and the PReNet denoiser are trained separately. For the 3D U-Net, 3D volume data were used for training. Due to limit of GPU memory, each 3D volume data was divided into patches of size 128×128×32. This U-Net was trained parallelly on 4 NVIDIA-GeForce GTX1080 Ti GPUs (11 GB memory). The batch size used for training was 28, the optimizer was the Adam optimizer with a learning rate of 5e-5, and the loss function was L1 loss. 560, 70 and 70 of the 700 simulated QSM images were used for training, validation, and testing, respectively. The output of the U-Net, the 3D volumes, were split into slices to train the 2D PReNet. The 2D patches all have patch size 128×128. The same optimizer, learning rate and loss function as for the 3D U-Net were used here. Parallel computing using the same GPU setting was adopted. A batch size of 60 was used to train the PReNet.

### 3.4 Evaluation of network performance

The qualitative improvement in motion‑corrupted and network‑corrected QSM volumes is assessed by visual inspection, and the quantitative improvement is examined by calculating root mean square error (RMSE), peak SNR (PSNR), and structural similarity index (SSIM) between motion‑corrupted, network‑corrected images and the reference image. PSNR approaches infinity and SSIM approaches 1 as the numerical difference between the images approaches 0. This assessment was performed on the image volumes with simulated motion artifact and real motion artifact. For the real motion artifact in PD patient images, the original and corrected images were only compared by visual inspection because the ground truth motion-free images are not available. For healthy subjects, however, the motion-corrupted and motion-correct images were registered to and compared with the ground truth QSM quantitatively. The registration was performed in MATLAB 2020a using the image processing toolbox.

## 4. Results

### 4.1 Qualitative evaluation: image results

The trained network achieved an overall improvement for the motion-corrupted testing images. Fig.3 shows the results for 1 representative subject with simulated motion. One slice of the axial view was plotted. From left to right are motion-free reference, motion corrupted QSM, motion-reduced QSM. The difference maps indicate a significant suppression of ringing and ghosting artifacts.

Fig.4 is the QSM for a representative healthy subject. The motion-corrupted (left) and motion-corrected (middle) are compared with the ground truth (right). Fig. 5 displays 1 representative axial QSM images from the PD patients. It can be seen that a significant amount of artifacts were suppressed, especially the ringing, leading to a visually better image quality for diagnosis.

### 4.2 Quantitative evaluation: image quality metrics

Boxplots of SSIM, PSNR and RMSE of the motion corrupted and motion corrected images are shown in Fig.6 for the validation set including 70 simulated examples. On each box, the central mark indicates the median, and the bottom and top edges of the box indicate the 25th and 75th percentiles, respectively. The whiskers extend to the most extreme data points not considered outliers. Quantitatively, the average SSIMs are 0.9006 and 0.9227 for motion-corrupted and motion-corrected dataset respectively, the average PSNR are 35.37 and 37.45 for motion-corrupted and motion-corrected dataset respectively, and the average RMSE are 0.0180 and 0.0141 (ppm) for motion-corrupted and motion-corrected dataset respectively. There is a substantial improvement in the image quality metrics and a significant decrease in the root mean square error.

Fig.7 displays the boxplots of SSIM, PSNR and RMSE of the motion corrupted and motion corrected images acquired from the 4 healthy subjects. Again, the central mark indicates the median, and the bottom and top edges of the box indicate the 25th and 75th percentiles, respectively. Quantitatively, the average SSIMs are 0.8167 and 0.8450 for motion-corrupted and motion-corrected images respectively, the average PSNR are 34.26 and 36.39 for motion-corrupted and motion-corrected images respectively, and the average RMSE are 0.0195 and 0.0152 (ppm) for motion-corrupted and motion-corrected images respectively. This result illustrates that the proposed network is effective in reducing real-world motion artifact.

## 5. Discussion

The preliminary results in simulated data, in healthy subjects, and Parkinson's patients presented in this paper show the ability of the proposed method to retrospectively correct motion artifacts in QSM. This represents the first attempt to correct motion artifacts in QSM using deep learning.

Unlike other MRI techniques where magnitude image matters for diagnosis purpose, QSM uses the phase image information of multi-echo GRE data to calculate susceptibility distribution of the brain. Strictly speaking, this phase image does not undergo the same affine motion transformation as the magnitude image but instead is equal to the dipole convolution of affine transformed susceptibility distribution, leading the known observation that the phase (or field) is orientation dependent (36). However, most real-world motions fall within a small range and, therefore the order of affine transformation and dipole convolution can be switched as an approximation. This avoids the requirement of computing a high-fidelity susceptibility map throughout the field of view when generating the training data for the proposed deep learning method. By combining the segments of k-space of different head states defined by a specific motion trajectory, a motion corrupted GRE signal is obtained by inverse Fourier transform. From this, a simulated motion corrupted QSM is computed using regular dipole inversion.

While the results show that motion artifacts are substantially reduced by the proposed method, some blurriness is introduced and some fine structures are smoothed. Motion artifacts tend to have various pattens depending on the position of motion events in the k-space and the magnitude of the motion. Motion occurring when the high spatial frequencies are sampled lead to ringing artifacts, while those occurring at low spatial frequencies often manifests as ringing and blurring. Even if the motion occur at the same position of the k-space, motion with larger magnitude or with different amounts of translation or ration will lead to changing artifacts. Therefore, when the neural network trying to adapt to a large variety of motion artifacts, it introduces blurring to decrease the loss function. In addition, the ghosting and blurring may induce misalignment of edges between motion-corrupted and reference images in each training pair. To compensate this misalignment the neural network blurs the image edges for a smaller loss function again.

We train our neural network with 56 brains and 10 simulated motion trajectories for each. Some of the brain and artifact features may be underrepresented by the training data, and the varied motion artifacts for each brain requires a neural network of sufficient complexity. Future directions include increasing the number of training examples in order to achieve a wider range of brain anatomy as well as artifact features. Furthermore, correcting motion artifact from the k-space is another potential direction. In this way, the source multi echo GRE data is corrected, and the corrected phase and magnitude images are used to reconstruct artifact-reduced using the MEDI routine, unlike in this work we correct artifacts directly on motion-corrupted QSM reconstructed from the motion-corrupted phase and magnitude images.

## 6. Conclusion

A deep neural network was trained to correct motion artifact in QSM, and it was tested on unseen data with synthetic motion artifacts and in vivo data with real motion-artifacts. Significant motion artifact suppression was observed.

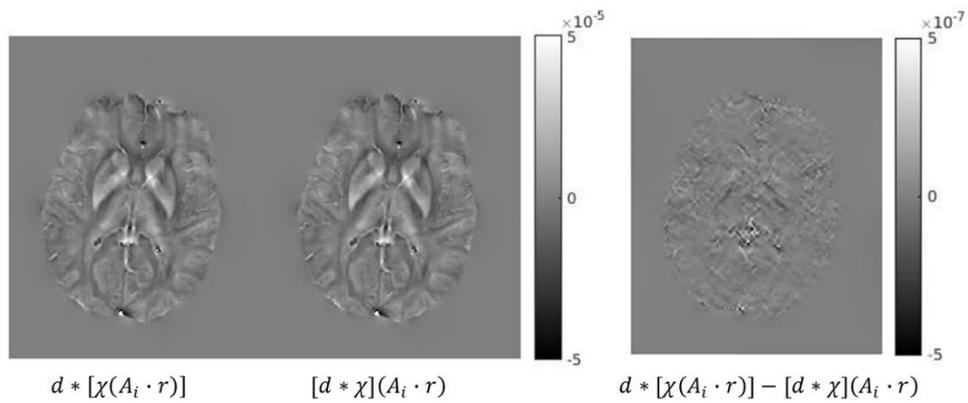

Figure 1: (a) Left: the field calculated by convolving the susceptibility and dipole kernel first and then rotating the field by 5° clockwise where $A_i$ stands for the affine matrix of the 5° clockwise rotation. Right: the field calculated by rotating the susceptibility by 5° clockwise and then convolving with the dipole kernel. (b) The difference map between the left and the right images in (a).

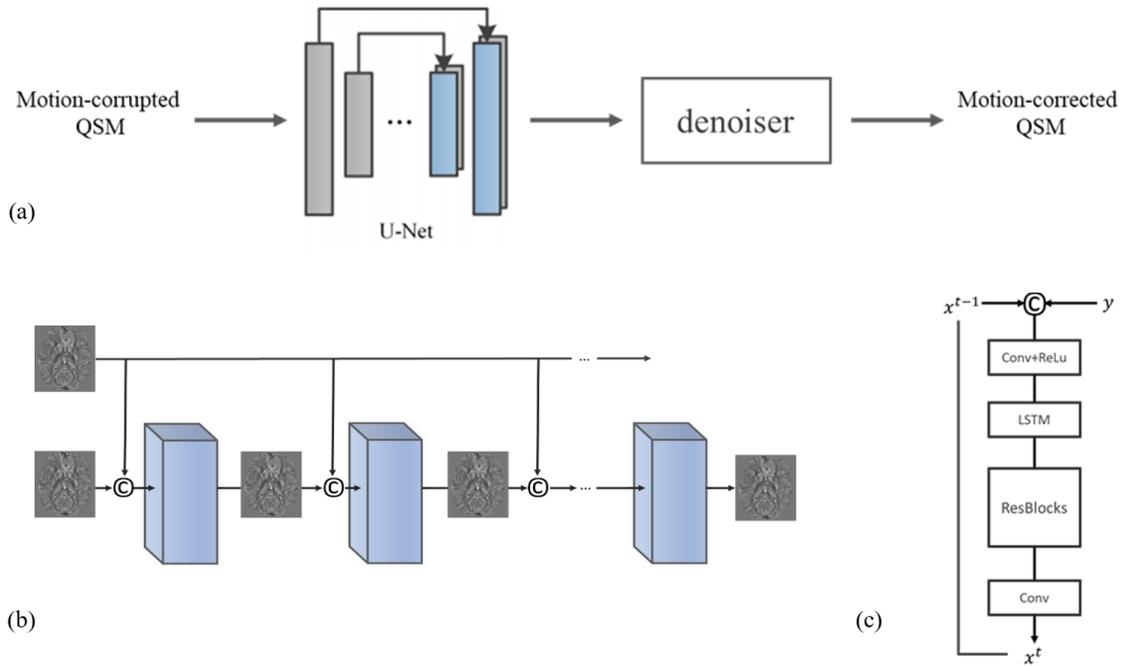

Figure 2. Network Architecture: (a) the flow of motion correction procedure: The motion corrupted image is passed through a 3D U-Net and then a 2D progressive recurrent neural network (PreNet). The U-Net and the PreNet are trained separately; (b) illustration of the PreNet with multiple-stage recursion: Ⓒ denotes concatenation; (c) the structure of the PreNet: It includes a convolution layer with ReLU, a convolutional LSTM, conventional ResBlocks and a final convolution layer.

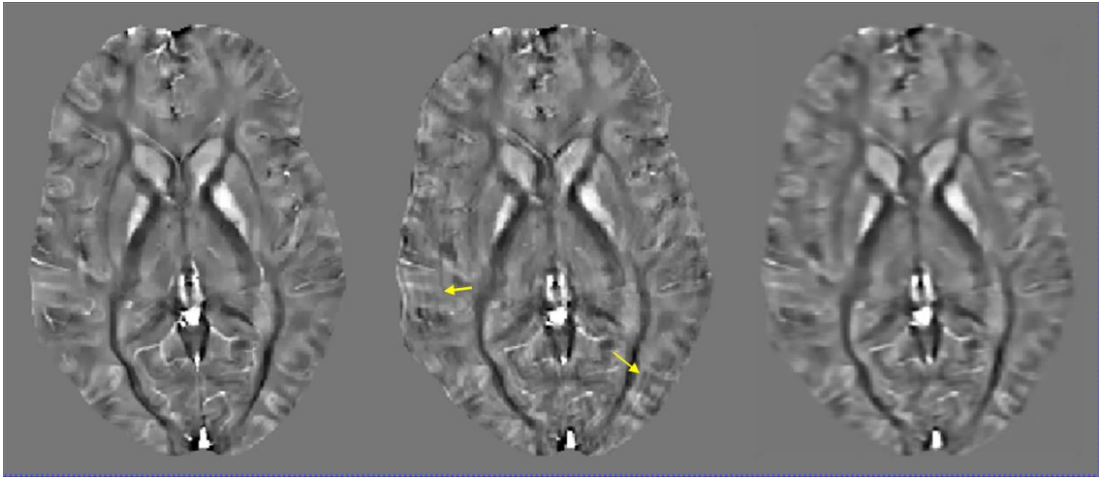

Figure 3: One representative example of simulated motion-corrupted QSM corrected by our neural network. From left to right are: motion-free reference, motion corrupted QSM, motion-reduced QSM. The yellow arrows point to the ghosting artifacts.

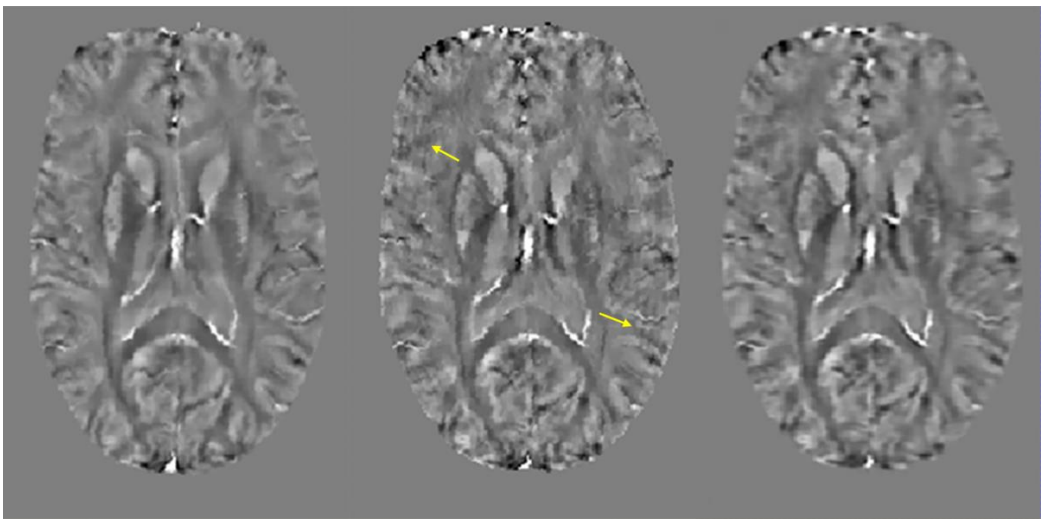

Figure 4: One representative example of a healthy volunteer. Two separate scans were performed. Left: The volunteer was asked to stay still throughout the scan. Middle: The volunteer was as asked to make 3 random and small movements during acquisition. Right: result of the motion-reduction applied to the motion-corrupted image.

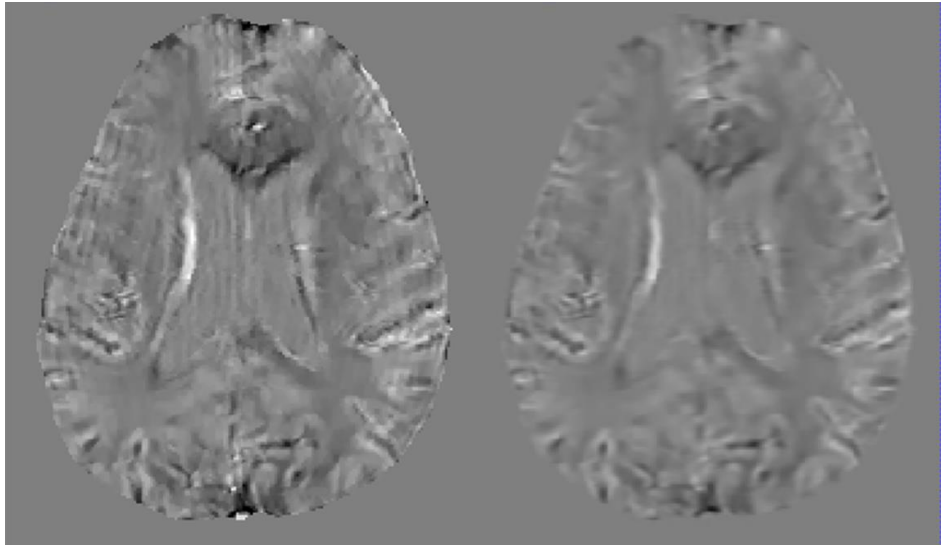

Figure 5: One representative example from Parkinson's disease patients with real motion artifacts where the ground truth motion-free reference is not available. The left displays the motion-corrupted image and the right displays the motion-reduced image corrected by our neural network.

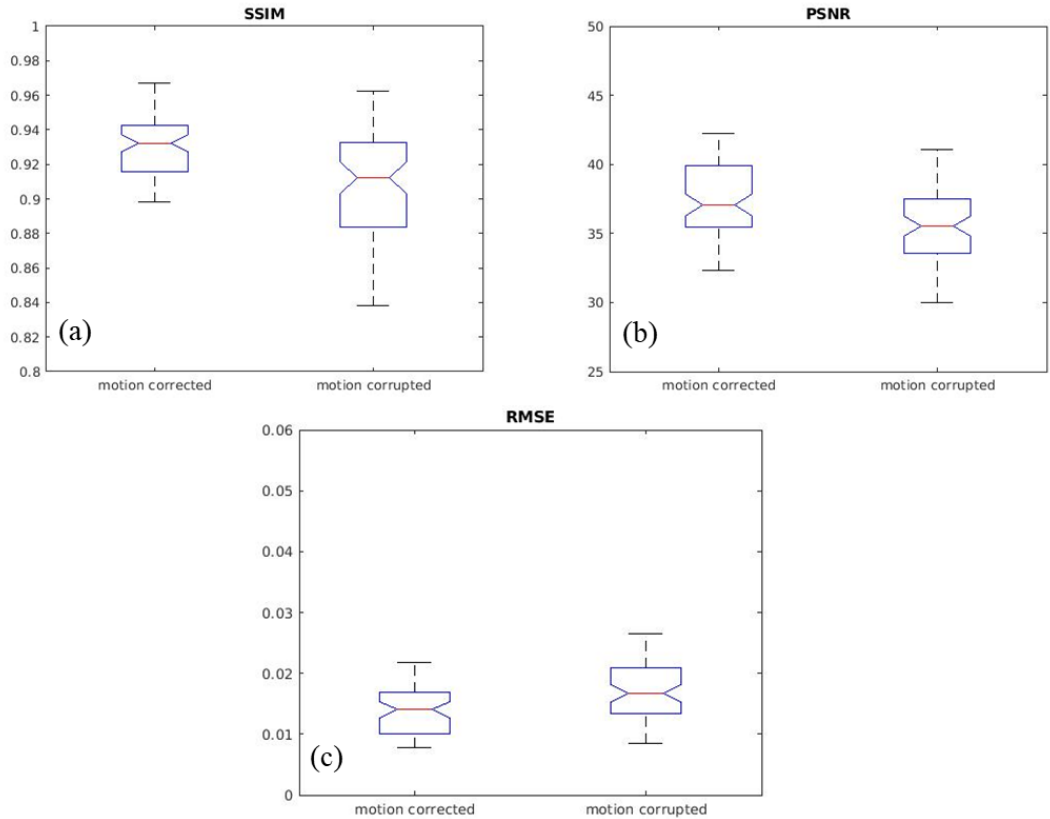

Figure 6. SSIM, PSNR and RMSE for the motion-corrupted versus motion-corrected image compared with the motion-free reference calculated on the simulated dataset: The top and the bottom of box represent 25$^{th}$ percentile and 75$^{th}$ percentile respectively; the red line represents the median.

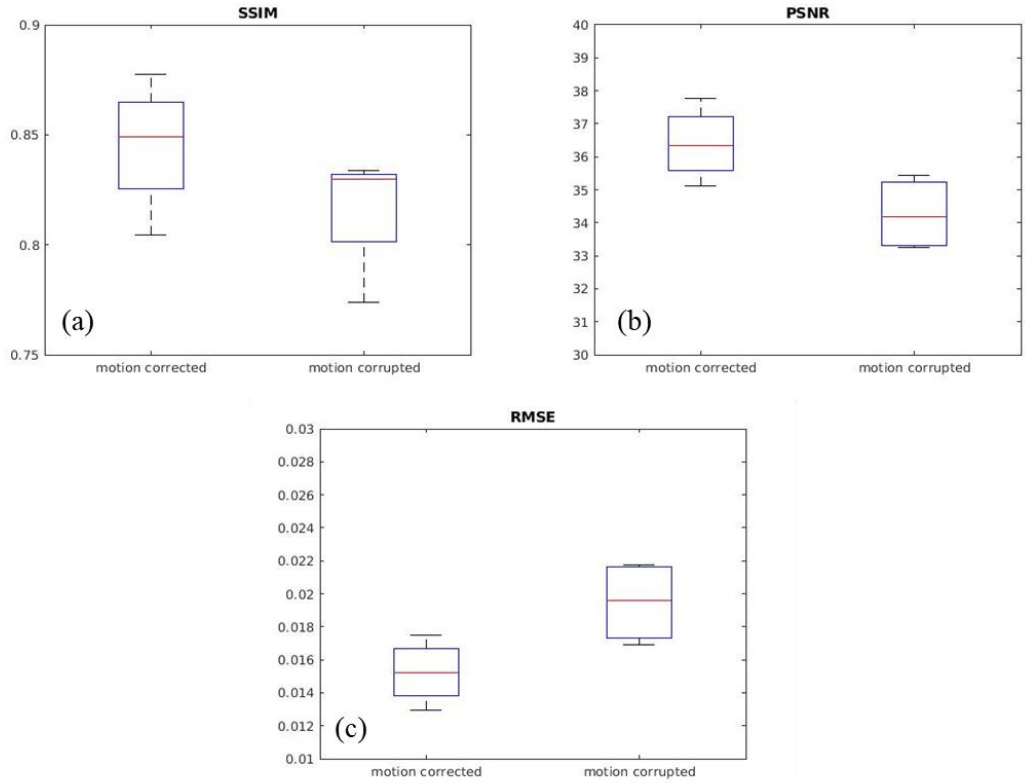

Figure 7. SSIM , PSNR and RMSE for the motion-corrupted versus motion-corrected image compared with the motion-free reference calculated on the dataset acquired from the 4 healthy volunteers which contained real-world motion artifacts: The top and the bottom of box represent 25th percentile and 75th percentile respectively; the red line represents the median.